\documentclass[prc,aps,tightenlines,floatfix,showpacs,twocolumn]{revtex4}
\usepackage[dvips]{graphics,color}
\usepackage{graphicx}
\usepackage{dcolumn}
\usepackage{bm}
\usepackage{epsfig}
\usepackage{supertabular}
\usepackage{amsmath}

\begin{document}
\def\ii{\'\i}

\title{
Phenomenological and microscopic cluster models I. The geometric 
mapping
%
%
%
}

\author{ H. Y\'epez-Mart\ii nez$^1$,
P. R. Fraser$^2$, P. O. Hess$^2$ and G. L\'evai$^3$  \\
{\small\it
$^1$Universidad Aut\'onoma de la Ciudad de M\'exico,
Prolongaci\'on San Isidro 151,} \\
{\small\it
Col. San Lorenzo Tezonco, Del. Iztapalapa,
09790 M\'exico D.F., Mexico} \\
{\small\it
$^2$ Instituto de Ciencias Nucleares, UNAM, Circuito Exterior, C.U.,} \\
{\small\it A.P. 70-543, 04510 M\'exico, D.F., Mexico} \\
{\small\it
$^3$Institute of Nuclear Research of the
Hungarian Academy of Sciences,} \\
{\small\it
Debrecen, Pf. 51, Hungary-4001} \\
}

\begin{abstract}
The geometrical mapping of algebraic nuclear cluster models is
investigated within the coherent state formalism. Two models are
considered: the {\it Semimicroscopic Algebraic Cluster Model} (SACM)
and the {\it Phenomenological Algebraic Cluster Model} (PACM), which
is a special limit of the SACM. The SACM strictly observes the Pauli
exclusion principle while the PACM does not.  The discussion of the
SACM is adapted to the coherent state formalism by introducing the new
$SO(3)$ dynamical symmetry limit and third-order interaction terms in
the Hamiltonian. The potential energy surface is constructed in both
models and it is found that the effects of the Pauli principle can be
simulated by higher-order interaction terms in the PACM. The present
study is also meant to serve as a starting point for investigating
phase transitions in the two algebraic cluster models.
\end{abstract}

\maketitle

\section{Introduction}
\label{one}

Phase transitions in algebraic models have been discussed since the
1970's \cite{gilmore}.  The principal method is the use of coherent
states \cite{kirson}.  The expectation value of the algebraic
Hamiltonian, with respect to a coherent state, is defined as the
semi-classical potential.  Through the behavior of that potential, as
a function in the parameter space, phase transitions and their order
can be studied.  The basic description, applied to the IBA \cite{iba},
was presented in \cite{kirson}. The method of coherent states also
delivers a geometrical mapping of the algebraic model in
consideration, providing in this manner an easy interpretation of the
dynamical symmetry limits.
Other studies on phase transitions in the IBA are published in
\cite{rowe-rosensteel}.

In \cite{cejnar}, a complete classification of phase transitions in
algebraic models is presented, restricted to Hamiltonians with up to
two-body interactions.  For the vibron models
\cite{roosmalen1,roosmalen2} the transitions turned out to be of
second order.  In \cite{zhang}, a second order phase transition was
also encountered.  In \cite{feng} the $U(3)$ boson model was studied
and it was noted that for very large number of bosons, the transition
may turn over into one of first order. The transitions were
investigated using the overlap of the ground state with that of the
$O(2)$ limit and searching for a step like behavior. 
In other words, 
no discontinuities of the derivatives of the potential were considered.

Coherent states have not only been applied to the IBA or atomic molecules
\cite{roosmalen1,roosmalen2}, but also to other algebraic models,
which have a microscopic origin within the shell model.  In
\cite{sympl} the geometrical mapping, using the vector coherent state
method \cite{hecht} was applied, mapping the pseudo-symplectic model
\cite{octavio-sympl} to the geometric model of the nucleus
\cite{eisenberg}.  The geometrical mapping turned out to be very
useful in calculations of nuclear spectra \cite{troltenier,appl-map}
and predicting the spectra of super heavy nuclei \cite{misicu}.  In
\cite{geom}, the coherent state method was used to obtain a
geometrical mapping of the {\it Semimicroscopic Algebraic Cluster
  model} (SACM)~\cite{sacm1,sacm2}. The SACM is an algebraic cluster
model which takes into account the Pauli exclusion principle.  It
allows for investigation of the effects of the Pauli exclusion
principle on the geometric potentials and on the order of the phase
transition.  Phase transitions in single nuclear systems were also
investigated in \cite{rowe-sym1,rowe-sym2}, related to the symplectic
model of the nucleus.

Note that phase transitions in nuclei, though not explicitly stated as
such, were already studied in 1972 in the first edition of
\cite{eisenberg}. A standard {\it curve discussion} was applied, while
a {\it phase transition} was denoted as a {\it shape transition}. In
the recent treatments, the main difference is the classification in
terms of the order of phase transitions.

Since its initial use, interest in the geometrical mapping, using the
coherent state method, has not been lost. One of the main reasons is
that multiple particle systems in conjunction with phase transitions
can be easily treated. These methods can provide important insight
into how to treat, in general, complicated many-body systems.

In recent years, phase transitions in atomic and nuclear molecules
were investigated in \cite{zhang,arias1,arias2,huitz1}, with the help
of the coherent state method.  The type of phase change discussed
there, important for the context of studies presented in this
contribution, is related to the $SU(3)$ and $SO(4)$ limit and the
transition between them in a molecule, which can consist of two atoms
or of two nuclei. Here we will restrict ourselves to nuclear clusters
only.  Two groups of models will be discussed: the 
{\it Phenomenological Algebraic Cluster Model} (PACM) and the {\it
Semimicroscopic Algebraic Cluster Model} (SACM). The {\it Vibron
Model} belongs to the group PACM \cite{vibron,who}.  In the PACM the
minimal number of relative oscillation quanta is always zero. In
contrast, in the SACM there is a minimal number of relative
oscillation quanta, $n_0$, due to the Wildermuth condition
\cite{wildermuth}, which is necessary in order to observe the
Pauli exclusion principle.  Though there is a lot of investigation in
the PACM on phase transitions, applications to actual nuclei are very
rare.  Only in \cite{bromley}, from the beginning of the vibron model,
and in \cite{barret} have we found applications to real cluster
systems.

Some of the important questions we would like to discuss in this and a
forthcoming publication are: What is the difference between taking
into account or not the Pauli exclusion principle?  What are the
orders of the phase transitions in the models discussed?  How does one
define the thermodynamical limit?  Normally, only second-order phase
transitions appear between the $SU(3)$ and $SO(4)$ dynamical
symmetries \cite{zhang}. So, is it also possible to obtain,
under certain circumstances, a first order phase transition?

This contribution restricts itself to the geometric mapping of an algebraic
Hamiltonian within the PACM and SACM. Already there some important
differences arise. One main result will be that in order that the PACM
reproduces the same results as the SACM, higher order interaction
terms are necessary which simulate the effects of the Pauli exclusion
principle. Differences and common features between the PACM and SACM
will be discussed.  The main reason for the differences is the large
overlap of the clusters, making it necessary to antisymmetrize the
many nucleon system. The PACM, which ignores the Pauli exclusion
principle, will consequently fail in satisfying basic conditions. For
atomic molecules this problem does not arise, because the two atomic
nuclei are separated in space, and thus no exchange effects play a
role. Some caution must be exercised using the comparison; because the
structures of the individual clusters are described within the SACM by
the shell model, we are obliged to compare to a PACM that also uses
the $SU(3)$ model. In general, the IBA model has been used in the
literature \cite{who,barret}. Additionally, the parameter by which the
number operator of $\pi$ bosons is multiplied is fixed in the SACM
because it describes the mean field. Within the PACM this parameter
can be chosen arbitrarily.

This all refers to the first question. The others will be addressed in
the second part, dedicated to the study of phase transitions.

The paper is structured as follows: In section \ref{two} the
Semimicroscopic Algebraic Cluster Model is revisited, introducing some
novel features, including the definition of the Phenomenological 
Algebraic Cluster Model (PACM) as a special limit of the SACM not
observing the Pauli principle.  In section \ref{three} the coherent
state formalism will be implemented in both models.  The PACM coherent
state can be recovered from the SACM when one sets the minimal number
of relative oscillation quanta, $n_0$, to zero.  In section \ref{four}
the geometrically mapped potentials are derived for the two models,
and finally, in section \ref{five} conclusions are drawn and a discussion
is presented on the differences of the PACM to the SACM.

\section{The Semimicroscopic Algebraic Cluster Model reconsidered}
\label{two}

In this section we present a brief overview of the Semimicroscopic
Algebraic Cluster Model\linebreak (SACM) \cite{sacm1,sacm2} and
introduce further amendments of it necessary for our study. These new
elements appear in all three subsections.
Previous applications of the SACM concerned describing the spectroscopic  
properties of core+$\alpha$-type 
\cite{c-und-a} and other 
\cite{other} two-cluster systems. 

\subsection{The group structure}
\label{two-one}

We start with reviewing the vibron model \cite{vibron}, in which  
the relevant degrees of freedom are oscillations in the relative 
motion of two structureless 
clusters in three dimensions. The operators describing them are
boson creation and annihilation operators with angular momentum one:
\begin{eqnarray}
\boldsymbol{\pi}_m^\dagger ~,~ \boldsymbol{\pi}_m ~~,~~ m=0,\pm 1 ~~~.
\end{eqnarray}
To this system one adds the spinless $\boldsymbol{\sigma}^\dagger$
boson creation and $\boldsymbol{\sigma}$ annihilation operators. They
define a cut-off, through the condition that the total number of
bosons $N=n_\pi + n_\sigma$ is kept constant. The $\sigma$-bosons have
no physical significance, which will play a role later on if one
intends to define a thermodynamical limit.  The $\boldsymbol{\pi}_m$
operators satisfy the relation
\begin{eqnarray}
\boldsymbol{\pi}^m & = & (-1)^{1-m}\boldsymbol{\pi}_{-m} ~~~.
\end{eqnarray}

The sixteen boson number conserving operators 
\begin{eqnarray}
\boldsymbol{\pi}_m^\dagger \boldsymbol{\pi}^{m^\prime} ~~,~~
\boldsymbol{\pi}_m^\dagger \boldsymbol{\sigma} ~~~,~~~
\boldsymbol{\sigma}^\dagger  \boldsymbol{\pi}^m ~~~,~~~ 
\boldsymbol{\sigma}^\dagger \boldsymbol{\sigma}
\end{eqnarray}
act as the generators of the $U_R(4)$ group, where $R$ stands for 
relative motion. There are two subgroup chains that contain the $SO_R(3)$ 
rotation group. The irreducible representations of the subgroups 
supply quantum numbers to define bases that are associated with the 
two dynamical symmetries: 
\begin{align}
&\hspace{0.3cm}U_R(4) \hspace{0.3cm} \supset \hspace{0.3cm} SU_R(3) 
\hspace{0.3cm} \supset \hspace{0.3cm} SO_R(3) \hspace{0.3cm} \supset
\hspace{0.3cm} SO_R(2)
\nonumber \\
&\lbrack N,0,0,0 \rbrack  \hspace{0.85cm}(n_{\pi} , 0 )   
\hspace{1.5cm}L_R  \hspace{1.7cm}M_R ,
\label{u3}
\end{align}
where
\begin{eqnarray}
n_{\pi} &=&  N, N-1,...,1,0,
\nonumber \\
L_R &=&  n_{\pi}, n_{\pi} -2,...,1 ~ \text{or} ~ 0,
\nonumber \\
M_R &=&  L_R, L_R-1, ...,-L_R ,
\label{u3part2}
\end{eqnarray}
and
\begin{align}
&\hspace{0.3cm}U_R(4) \hspace{0.3cm} \supset \hspace{0.3cm} SO_R(4) 
\hspace{0.3cm} \supset \hspace{0.3cm} SO_R(3) \hspace{0.3cm} \supset 
\hspace{0.3cm} SO_R(2)
\nonumber \\
& \lbrack N,0,0,0 \rbrack  \hspace{1cm} (\omega ,0)
\hspace{1.6cm}L_R  \hspace{1.7cm}M_R ,
\label{o4}
\end{align}
where
\begin{eqnarray}
\omega &=& N, N-2,...,1~ \text{or} ~0,
\nonumber \\
L_R &=& \omega, \omega -1,...,1, 0,
\nonumber \\
M_R &=& L_R, L_R-1, ...,-L_R .
\label{o4part2}
\end{eqnarray}
The $SU(3)$ dynamical symmetry {\it is generally believed
to be the vibrational limit}
of the system around a spherical equilibrium shape, while
the $SO(4)$ dynamical symmetry describes static dipole deformation.

The vibron model formalism reviewed up to this point handles only the 
relative motion of 
the clusters and neglects their internal sructure. In order to 
incorporate these degrees of freedom too, the SACM applies  
Elliott's $SU(3)$ model \cite{elliott}. The orbital structure of 
the clusters is then described by the $SU_{C_k}(3)$ group, where 
$C_k$ refers to  the $k^{\text{th}}$ cluster, $k=1,\ 2$. The Elliott model 
applies $LS$ coupling, but in many cases the $S$ spin degree of freedom 
does not play a role. This is the case, for example with even-even 
clusters, and for the sake of simplicity we shall consider clusters 
of this type in what follows. 

It is essential that in the SACM the $SU(3)$ group appears not only in
the description of the relative motion and the individual clusters,
but also in the description of the unified nucleus. The typical group
structure associated with a two-cluster system in the SACM is then
\begin{align}
& SU_{C_1}(3) \otimes SU_{C_2}(3) \otimes SU_R(3) \:\: \supset \:\: 
SU_C(3) \otimes SU_R(3) \:\: \supset
\nonumber \\
& \; (\lambda_1,\mu_1) \hspace{0.55cm} (\lambda_2,\mu_2) \hspace{0.5cm}
(n_\pi,0) \hspace{1.6cm} (\lambda_C,\mu_C)
\nonumber \\
& \hspace{1.6cm} SU(3) \hspace{0.3cm} \supset \hspace{0.3cm} SO(3)
\hspace{0.3cm} \supset \hspace{0.3cm} SO(2)
\nonumber \\
& \hspace{1.65cm} (\lambda , \mu )  \hspace{1.4cm} \kappa L \hspace{1.6cm} M,
\label{chain}
\end{align}
where $(\lambda_k,\mu_k)$ refers to the $SU_{C_k}(3)$ irreducible
representations (irreps) of the individual clusters, which are then
coupled to intermediate irrep $(\lambda_C,\mu_C)$. These irreps are
the ones associated with the ground-state configuration of the $k^{\text{th}}$
cluster.  $n_\pi$ is the number of relative oscillator quanta, while
$(\lambda , \mu )$ is the total $SU(3)$ irrep. $L$ and $M$ are the
angular momentum and its projection, and $\kappa$ is used to
distinguish multiple occurrences of a given $L$ in $ (\lambda , \mu)
$.

The model space of the SACM is obtained by comparing all possible
irreps $(\lambda ,\mu)$, as given in (\ref{chain}), contained in the
product $(\lambda_{C_1},\mu_{C_1})$ ${\otimes}$
$(\lambda_{C_2},\mu_{C_2})$ ${\otimes}$ $(n_\pi ,0)$, with those of
the shell model and retaining only those irreps which appear in the
shell model.  Computer codes determining the model space are available
and can be obtained on request. In most cases, however, it is easy to
retrieve the irreps by hand.  In this manner the Pauli exclusion
principle is observed (for some illustrative examples, see
Refs.~\cite{sacm1,sacm2}).  The $SU(3)$ basis is also useful in
eliminating the spurious center of mass motion.

We note that the above $SU(3)$ matching procedure also reproduces the
Wildermuth condition \cite{wildermuth} in a natural way.  This
condition prescribes a minimal number of oscillator quanta
(i.e. $n_{\pi}$) in the relative motion. Apart from the case of
closed-shell clusters, however, it is only a necessary condition for
the handling of the Pauli exclusion principle.

It is now worthwhile to discuss the possible dynamical symmetries of
the SACM based on those of the vibron model. The $SU(3)$ dynamical
symmetry is clearly associated with the (\ref{chain}) group chain. The
equivalent of the $SO(4)$ dynamical symmetry of the vibron model,
however, can be considered only an approximate dynamical symmetry in
the SACM. The reason is that due to the Pauli principle part of the
set of $SO(4)$ basis states has to be excluded from the model
space. Although $n_{\pi}$ is not a good quantum number in the $SO(4)$
limit, the $SO(4)$ basis states can be written as linear combinations
of $SU(3)$ states, so excluding these below the minimal allowed
$n_{\pi}$ value distorts the structure of the $SO(4)$ basis. Finally,
a third dynamical symmetry can also be derived from the $SU(3)$
dynamical symmetry of the vibron model.  The group structure
associated with this $SO(3)$ dynamical symmetry is
\begin{align}
& SU_C(3) \otimes U_R(4) \supset  SO_C(3) \otimes SO_R(3) \supset
 SO(3) \supset SO(2)
\nonumber \\
&(\lambda_C.\mu_C)\, \lbrack N,0,0,0 \rbrack ~~~~ L_C ~~~~~~~~~~ L_R ~~~~~~~~~~ 
L ~~~~~~~~~  M .
\nonumber \\
\label{o3}
\end{align}

The difference between the previously mostly ignored (\ref{o3}) chain
and the one appearing in (\ref{chain}) is of dynamical nature in the
sense that the interaction in the former case does not contain terms
typical of the coupled $SU(3)$ degrees of freedom. In fact, the
$SU(3)$ groups do not play a role other than supplying labels for
classification of the states. In terms of interactions we can call the
scenarios associated with the (\ref{o3}) and (\ref{chain}) chains as
weak and strong coupling limits, respectively. The two limits are the
same when the two clusters are both closed-shell nuclei, but when at
least one of them is not (i.e. its internal $(\lambda_k,\mu_k)$ irrep
is different from (0,0)), a clear difference between the two limits
arises.

Before closing this subsection it is worthwhile to comment on the
typical selection rules characterizing the dynamical symmetries. This
is also related to the band structure determined by the appropriate
group structure. In the basis associated with the $SU(3)$ dynamical
symmetry of the SACM the bands are defined by the $(\lambda,\mu)$ and
$\kappa$ quantum numbers (see Eq. (\ref{chain})), where $\kappa$ is
obsolete when either $\lambda$ or $\mu$ is zero, as is the case in the
$SU(3)$ limit of the vibron model too (see Eq. (\ref{u3})). The states
belonging to the same $SU(3)$ irrep are connected by the quadrupole
operator, the $SU(3)$ tensorial character of which is $(1,1)$. This
operator leaves $n_{\pi}$ and the parity intact and changes the
angular momentum by two units, so it describes electric quadrupole
transitions.  On the other hand, bands associated with the $SO(4)$
dynamical symmetry are characterized by the $\omega$ quantum number
defining the $SO(4)$ irreps (see Eq. (\ref{o4})) and contain states
with both even and odd angular momentum, i.e. with both positive and
negative parity.  The in-band transitions are described by the $SO(4)$
generators, which play the role of the electric dipole operator. The
two dynamical symmetries thus lead to different selection rules, and
this has to be taken into account when they are applied to some
concrete physical problem.

\subsection{The Hamiltonian}
\label{two-two}

Let us now turn to the Hamitonian associated with the SACM. While in
most typical applications it is sufficient to consider interaction
terms constructed as two-body terms, here we argue that a specific
third-order interaction term is also necessary to stabilize the
spectrum.  Furthermore, as another new element we shall separate the
Hamiltonian into terms associated with the three dynamical symmetries
identified above. The parametrization introduced this way allows
interpolation between the dynamical symmetries, changing certain parameters
of the Hamiltonian, like $x$ and $y$ (see equation below).
We consider two cases: i) both clusters spherical and ii)
one spherical cluster plus a deformed one.  Examples for these two
scenarios are the $^{16}$O+$\alpha$ $\rightarrow$ $^{20}$Ne and
$^{20}$Ne+$\alpha$~$\rightarrow$~$^{24}$Mg systems, examined in Paper-II
of this series.

The Hamiltonian is given by
\begin{equation}
\boldsymbol{H}  =  xy\boldsymbol{H}_{SU(3)} 
+ y(1-x)\boldsymbol{H}_{SO(4)}+(1-y)\boldsymbol{H}_{O(3)}
\label{H-tot}
\end{equation}
with
\begin{eqnarray}
\boldsymbol{H}_{SU(3)} & = &
\hbar \omega \mbox{\boldmath$n$}_{\pi }+a_{Clus}
\mathit{C}_{2}\left( \lambda _{C},\mu _{C}\right)
\nonumber \\
&&
+(a-b\Delta
\mbox{\boldmath$n$}_{\pi })\mathit{C}_{2}\left( \lambda ,\mu \right)  
+(\bar{a}-\bar{b}\Delta
\mbox{\boldmath$n$}_{\pi })\mathit{C}_{2}\left( \boldsymbol{n}_\pi ,0 \right) 
\nonumber \\
&&
+\gamma {\mbox{\boldmath$L$}}^{2}+t{\mbox{\boldmath$K$}}^{2}
\nonumber \\
\boldsymbol{H}_{SO(4)} & = & 
a_{C}{\boldsymbol{L_{C}}}^{2}
+a_{R}^{\left( 1\right) }{\boldsymbol{L_{R}}}%
^{2}
\nonumber \\
&&
+\gamma {\boldsymbol{L}}^{2}
+\frac{c}{4}\left[ (\mbox{\boldmath$\pi$}^{\dagger }\cdot %
\mbox{\boldmath$\pi$}^{\dagger })-(\sigma ^{\dagger })^{2}\right] \left[ (%
\mbox{\boldmath$\pi$}\cdot \mbox{\boldmath$\pi$})-(\sigma )^{2}\right] 
\nonumber \\
\boldsymbol{H}_{O(3)} & = &
\hbar \omega \mbox{\boldmath$n$}_{\pi }
\nonumber \\
&&+a_{C}{\boldsymbol{L_{C}}}^{2}+a_{R}^{\left( 1\right) }{\boldsymbol{L}}%
_{R}^{2}+\gamma {\mbox{\boldmath$L$}}^{2}
~~~,
\label{one-d}
\end{eqnarray}
with $\Delta \boldsymbol{n}_\pi = \boldsymbol{n}_\pi - n_0$, $n_0$
being the minimal number of quanta.
The $a_{Clus}$ is the strength of the quadrupole-quadrupole
interaction, restricted to the cluster part, while $R$ and $C$
denote the contributions related to the {\it relative} and
{\it cluster} part respectively.
Further interaction terms are the
total angular momentum operator, $\boldsymbol{L}^2$, and the
$\boldsymbol{K}^2$ operator, defined in \cite{sacm1,sacm2} which
classifies the rotational bands, giving the projection of the angular
momentum onto the intrinsic $z$ axis.  For the case of two spherical
clusters, the second-order Casimir operator of $SU(3)$ is just given
by $\boldsymbol{n}_\pi ( \boldsymbol{n}_\pi + 3)$.  Note that in the
case of deformed clusters the information about the deformation only
enters in the $SU(3)$ dynamical limit.

Note that the division in (\ref{one-d}) is done according to
dynamical symmetry limits and not according to two terms, referring to
each cluster, one to the relative motion and one to the interactions
between them. If one wishes to do that, all what has to be done
is to decouple the different contributions. For example, the
$\boldsymbol{L}^2$ operator can be written as $\left( \boldsymbol{L_C}
+ \boldsymbol{L}_R \right)^2$ = $\left[ \boldsymbol{L}_C^2 +
  \boldsymbol{L}_R^2 + 2 \left( \boldsymbol{L}_C \cdot
  \boldsymbol{L}_R \right)\right]$. The first and second term refer to
the cluster and relative angular momentum, respectively, while the
last term refers to the coupling between the channels. This can be
further divided by writing the cluster angular momentum as\linebreak
$\boldsymbol{L}_{Cm}= \left( \boldsymbol{L}_{C_1m} + \boldsymbol{L}_{C_2m}
\right)$.

The division according to dynamical symmetries in (\ref{one-d}) was
done as follows: In the $SU(3)$ limit the coupling of interaction
operators is on the level of $SU(3)$, i.e., Casimir operators of
$SU_C(3)$ and $SU(3)$ have to appear. This is called the {\it strong
  coupling limit}.  The $SO(3)$ limit couples only at the $SO(3)$
level, i.e. only the angular momentum operators appear (no second- and
higher-order $SU(3)$ Casimir operators, except $\boldsymbol{n}_\pi$
and functions in it). This is called the {\it weak coupling limit}.
This is reflected by the appearance of interaction terms related only
to angular momentum ($\boldsymbol{L}_R^2$, $\boldsymbol{L}_C^2$ and
$\boldsymbol{L}^2$).  In the literature one usually refers to this
latter limit as the $SU(3)$ limit, where the Hamiltonian contains only
the $\hbar\omega \boldsymbol{n}_{\pi}$ term plus at most some weak
anharmonic terms.  Here we feel it necessary to change the notation
because we understand the $SU(3)$ limit to include also terms such as
a strong quadrupole-quadrupole interaction. In order to compare with
results in the literature, this has to be kept in mind when we report
on phase transitions between different dynamical symmetries.  The
$SO(4)$ limit is defined through the appearance of the second-order
Casimir operator of $SO(4)$. This limit is called the {\it deformed
  limit} because the interaction will always produce a potential with
a deformed minimum.

In principle, one can add the angular momentum operator of the
deformed clusters ($\boldsymbol{L}_k^2$, $k=1,2$). We exclude this
interaction for the moment.

The new higher-order interaction appearing in the third term of
$\boldsymbol{H}_{SU(3)}$ needs some explanation.  The whole term is
related to the quadrupole-quadrupole interaction, {\it which is
  present in any nuclear system}.  However, without the
$-b\Delta\boldsymbol{n}_\pi$ (with $-b>0$) correction, states which
contain a sufficiently large $n_\pi$ will be lower in energy than
states with the minimal number of $\boldsymbol{\pi}$ bosons, $n_0$.
This is due to the dependence on $\boldsymbol{n}_\pi^2$ in the
second-order Casimir operator, which will finally dominate over the
$\hbar\omega\boldsymbol{n_\pi}$ term for a sufficiently large number
of $\boldsymbol{\pi}$-bosons.  In the standard treatment, when $n_\pi$
is conserved, a simple restriction to small $\Delta n_\pi$ suffices to
circumvent the problem, i.e., states with large $\Delta n_\pi$ are
simply not taken into account in the model space.

This effect was studied in \cite{TE1} within the context of the
symplectic model of the nucleus \cite{octavio-sympl,rowe1,rowe2}.
Also there, the quadrupole-quadrupole interaction dominates over the
kinetic energy and finally will promote high $n_\pi$ states to low
energies, even below the physical ground state. This problem was
solved by subtracting from the quadrupole-quadrupole interaction the
so called {\it Trace Equivalent} part \cite{TE1}, which insures that
the average mean field is still represented by a harmonic potential.
When no correction is applied, the mean field shell structure is
destroyed and a mean oscillator structure, one of the main assumptions
of the shell model, cannot be assumed anymore.  This was also noted
within the SACM in \cite{sacm2}, where correction terms of the type
$\Delta n_\pi$, mentioned here, were included.  Without these
corrections the problem increases significantly when interactions
mixing states with different $n_{\pi}$ are considered.  Then, avoiding
states with large $n_\pi$ is not an option, as is in the case of
conserved $n_\pi$, when the model space can be limited in $n_\pi$
using physical arguments.

\subsection{PACM: the phenomenological limit of the SACM}
\label{two-three}

The minimal number of $\pi$ bosons is an essential requirement in 
the SACM to incorporate the Pauli principle. However, the formalism 
allows setting this minimal number to zero. This limit of the SACM 
can be defined as the Phenomenological Algebraic Cluster Model (PACM). 
It has to be stressed that the difference between the SACM and the 
PACM manifests itself only in the model space, while the two models 
share the same Hamiltonian and other operators. Obviously, the 
different model space will lead to different matrix elements in 
the two models.
Note that the minimal number of relative oscillation quanta is
either 0 or $n_0$. It is not allowed to choose a number
in between, because each of such a number violates the Pauli exclusion
principle.

When both clusters are closed-shell nuclei, the PACM essentially 
recovers the vibron model \cite{vibron}. 

One of the main objectives of the present work is to investigate 
the similarities and differences between the two approaches. This 
is especially interesting within the context of the coherents state 
formalism, because in other models restrictions similar to those 
in the SACM (i.e. restricting the boson number) are unknown. In this 
sense the formalism of the PACM is closer to that of other models. 
Due to the minimal number of $\pi$ bosons the formalism of the SACM 
will obviously become more involved. It is our aim to explore this 
conflict between the physical importance of a fundamental principle 
(i.e. the Pauli principle) and the technically more complicated 
formalism arising due to it.

\section{Coherent states and the geometrical mapping}
\label{three}

In this section the coherent state is presented, which is used to
obtain a geometrical mapping of the SACM and PACM in the next section.

The use of coherent states is the most common method of applying a
geometrical mapping
\cite{gilmore,kirson,roosmalen1,roosmalen2,sympl,geom}.  One advantage
is that the coherent state can be expanded in terms of the complete
set of states for a given total number of bosons, $N$,
(in the SACM, this refers to all allowed
basis states for a given total number of quanta).  The ground-state
energy is usually reproduced very well. The coherent state also
provides a transparent relation to collective variables.  Its use is
justified by noting that it corresponds to the {\it Gaussian Overlap
  Approximation} within the {\it Generator Coordinate Method}
\cite{ring}, skipping the term of the zero-point motion. As shown in
\cite{ring}, this method allows the definition of a potential with
usually good results.  However, the mass parameters of the kinetic
energy are usually not reproduced very well. In order to obtain a
kinetic energy too, the coherent state variables have to be defined as
complex variables \cite{roosmalen1,roosmalen2}.  We do not consider
the kinetic energy due to the reason mentioned above, and focus on the
potential.

The coherent state within the SACM was introduced in \cite{geom}
\begin{eqnarray}
\left| \alpha \right> &=& 
{\cal N}_{N,n_0}
(\boldsymbol{\alpha} \cdot \boldsymbol{\pi})^{n_0}
\left[
\boldsymbol{\sigma}^\dagger 
+ \left( \boldsymbol{\alpha} \cdot \boldsymbol{\pi}^\dagger
\right)\right]^{N} \left| 0 \right>
\nonumber \\
&=& {\cal N}_{N,n_0} \frac{N!}{(N+n_0)!}
\frac{{\rm d}^{n_0}}{{\rm d}\gamma^{n_0}_1}
\left[
\boldsymbol{\sigma}^\dagger 
+ \gamma_1 \left(\boldsymbol{\alpha} \cdot \boldsymbol{\pi}^\dagger
\right)\right]^{N+n_0} \left| 0 \right> \, ,
\nonumber \\ 
\label{coh-wp}
\end{eqnarray}
where, for convenience, we redefined the total number of relative
oscillation quanta as $(N+n_0)$, while the $\gamma_1$ parameter has to
be set equal to 1 after the differentiation.

The normalization factor is given by \cite{geom}
\begin{equation}
{\cal N}^{-2}_{Nn_0} =
\frac{N!^{2}}{(N+n_{0})!}\frac{{\rm d}^{n_{0}}}{%
{\rm d}\gamma_{1}^{n_{0}}}\frac{{\rm d}^{n_{0}}}{
{\rm d}\gamma_{2}^{n_{0}}}\left[%
1+\gamma_{1}\gamma_{2}(\boldsymbol{\alpha}^{*}\cdot 
\boldsymbol{\alpha})\right]^{N+n_{0}}
\;.
\end{equation}
Again, the $\gamma_k$ have to be set equal to 1 after the application
of the derivatives.

The $\boldsymbol{\alpha}$ is a short-hand notation for the, in
general, complex variables $\alpha_{m}$ ($m=1,0,-1$).  The coherent
state with complex $\boldsymbol{\alpha}$ coefficients is the most
general linear combination of the boson creation operators.  For
static problems the requirement \cite{roosmalen1,roosmalen2}
\begin{eqnarray}
\alpha_{m}^{*}  & = & (-1)^{1-m} \alpha_{-m} ,
\end{eqnarray}
reduces the number of real parameters to three, namely to $\alpha_0$
and the real plus the imaginary part of $\alpha_{+1}$.

In the Appendix we present the results for the geometrical mapping
for the important interaction terms appearing in the Hamiltonians.
We define
\begin{eqnarray}
(\boldsymbol{\alpha} \cdot \boldsymbol{\alpha}) & = &
\sum_m (-1)^{1-m} \alpha_m \alpha_{-m}
\nonumber \\
& = & \alpha^2
~~~,
\end{eqnarray}
where $\alpha$ represents a measure of the inter-cluster distance
\cite{geom} and
$\alpha^2$ a short-hand notation for $(\boldsymbol{\alpha} \cdot
\boldsymbol{\alpha})$.  Because the only relevant variable is the
inter-cluster distance, we can express the potential in terms of this
sole variable $\alpha$.

The importance of the coherent states resides in the fact that they
provide us with the possibility to define a {\it Potential Energy
Surface} (PES)
\begin{eqnarray}
V(\boldsymbol{\alpha}) & = &
\left< \boldsymbol{\alpha} \right| \boldsymbol{H} 
\left| \boldsymbol{\alpha} \right>
~~~, 
\end{eqnarray}
in terms of  the collective variables $\alpha_m$.

\subsection{Renormalization of the variable $\alpha$} 
\label{three-one}

It is often convenient to transform $\alpha$ to other related
variables.  In the literature there are different {\it conventions},
which often contradict each other\linebreak
\cite{cejnar,roosmalen1,roosmalen2,zhang,arias1,arias2,huitz1,lemus1,lemus2,
levit}. Here we present some arguments to justify our choice, restricting 
for simplicity to the PACM, where $n_0=0$.

In \cite{roosmalen1,roosmalen2} the renormalization of the interaction
parameters is proposed using the following reasoning: the expectation
value of a one-body interaction with respect to the coherent state is
proportional to the total number of quanta.  As an example, the
expectation value of $a_1 \boldsymbol{n}_\pi$ is given by
\begin{eqnarray}
\langle a_1 \boldsymbol{n}_\pi \rangle & = & a_1 N \frac{\alpha^2}
{\left( 1+\alpha^2\right)} ~~~.
\label{an}
\end{eqnarray}
Assuming that $\alpha$ is of the order of one, this expectation value
increases with $N$, which is unnatural because $\langle
\boldsymbol{n}_\pi \rangle$ is of the order of one (note that this
argument {\it assumes} that $\alpha$ is of the order of one, too).  In
order to avoid this, it is recommended in \cite{roosmalen1,roosmalen2}
to redefine the interaction as
\begin{eqnarray}
\frac{a_1^\prime}{N} \boldsymbol{n}_\pi ~~~.
\label{1-body}
\end{eqnarray}
In this manner, it is expected that the parameter $a_1^\prime$ does
not change significantly with increasing $N$.  A similar argument
holds for the two-body interaction, recommending to divide the
corresponding interaction parameter by $N(N-1)$, and so on.

This argument is supported by considerations given in \cite{jolie}.
There it is shown that one- and two-body interactions scale like $N$
and $N(N-1)$, respectively.  This dependence has to be canceled in
order to define a thermodynamic limit $N\rightarrow\infty$. However,
the $N$ used in \cite{jolie} can always be related to the number of
particles. For example, in the IBA \cite{iba} $N$ is given by half the
number of valence nucleons and in Lipkin-type models (two-level
systems) the $N$ is given by the number of states in the lower level
which are completely filled in its lowest states. In the algebraic
cluster models, however, the $N$ is {\it not related to a number of
  particles} but rather to a boson {\it cut-off}.  The $\sigma$-boson is
introduced merely to define a cut-off for the number of $\pi$-bosons
($n_\pi \leq N$). Physics requires that the final results do not
depend on this cut-off.

In what follows we present a proposal on how to treat the cut-off in
algebraic cluster models.  Note that in (\ref{1-body}) a conceptual
problem arises already, which has not been considered before in the
literature.  Here we try to convince the reader that the former
renormalization (simply dividing the interaction parameters by a power
in $N$) is too simple and a more sophisticated renormalization has to
be applied.  Here we illustrate the situation in three different ways.
\\
a) Consider first the $\hbar\omega \boldsymbol{n}_\pi$ term, which
represents the mean field of the harmonic oscillator in the
SACM. Dividing it by $N$ implies that $\hbar\omega^\prime =
\frac{\hbar\omega}{N}$ tends to zero as $N\rightarrow\infty$.  The
contradiction becomes apparent because $\hbar\omega$ is a physical
value, which has to be kept independent of $N$, while $a_1$ in
Eq. (\ref{an}) is used as an abstract parameter.
\\
b) Looking at it from a different angle, let us consider the operator
$\boldsymbol{n}_\pi$ alone.  When applying $\boldsymbol{n}_\pi$ in an 
harmonic oscillator basis, in the lowest states the eigenvalues of
$\boldsymbol{n}_\pi$ are small numbers.  The geometrical mapping gives
$N\frac{\alpha^2}{\left( 1+\alpha^2\right)}$, which suggests an
increase proportional to $N$.  The only way to maintain numbers of the
order of one is to {\it redefine} $\alpha$ by
\begin{eqnarray}
\alpha^2 & \sim & \frac{\delta^2}{N} ~~~,
\end{eqnarray}
such that the $\alpha$ tends to zero for $N\rightarrow\infty$.  In
this way, with $\delta$ of the order of one, the geometrical mapping
also gives results of the order of one. Thus,
restricting for example to the PACM ($n_0=0$), the coherent state in
(\ref{coh-wp}) is redefined as
\begin{equation}
\left| \boldsymbol{\delta} \right> =
\frac{1}{\sqrt{N!(1+\frac{1}{N}\left[\boldsymbol{\delta}
\cdot\boldsymbol{\delta} \right])^N}}
\left[
\boldsymbol{\sigma}^\dagger + \frac{1}{\sqrt{N}}
\left( \boldsymbol{\delta} \cdot \boldsymbol{\pi}^\dagger
\right)\right]^N \left| 0 \right> \,.
\label{coh-wop-2}
\end{equation}
The expectation value of $\boldsymbol{n}_\pi$ then leads to
\begin{eqnarray}
\langle \boldsymbol{n}_\pi \rangle & = & \frac{\delta^2}
{\left( 1+\frac{1}{N}\delta^2\right)} ~~~.
\end{eqnarray}
The global dependence on $N$ {\it vanishes}, while for
$N\rightarrow\infty$, the $N$-dependence disappears.

The same happens for most of the two-body interactions, as we will
discuss further below, with some differences when the
$\boldsymbol{\sigma}$~operators are involved.
\\
c) The above choice of the new variable $\delta$ can be justified in a
{\it third way}. To prove this, let us consider the coordinate
operator defined without $\sigma$-bosons, i.e. without cut-off ($N$),
namely
\begin{eqnarray}
\boldsymbol{r}_m & = & \sqrt{ \frac{\hbar}{2m\omega}}
\left( \boldsymbol{\pi}^\dagger_m + \boldsymbol{\pi}_m \right)
~~~.
\label{def-r-1}
\end{eqnarray}
When the $\sigma$-bosons are introduced, this operator has to be
changed. The new operator, called the {\it algebraic coordinate
operator}, should satisfy the following minimal conditions: i) The
total number of bosons has to be kept constant, i.e., each
$\boldsymbol{\pi}_m^\dagger$ has to be multiplied by
$\boldsymbol{\sigma}$ and each $\boldsymbol{\pi}_m$ has to be
multiplied by a $\boldsymbol{\sigma}^\dagger$; ii) the definition of
the distance operator should be {\it independent} of the basis and
Hamiltonian used; and iii) for $N\rightarrow\infty$ it should converge
to the standard form given in (\ref{def-r-1}).  The proposed algebraic
coordinate operator is given by
\begin{equation}
\boldsymbol{r}_m^a  = \sqrt{ \frac{\hbar}{2Nm\omega}}
\left( \boldsymbol{\pi}^\dagger_m \boldsymbol{\sigma} +
\boldsymbol{\sigma}^\dagger \boldsymbol{\pi}_m \right)
~~~.
\end{equation}
The ``$a$" refers to an {\it algebraic} operator. The operator
itself does not change the total number of bosons, as required by the
above condition i). The $N$ in the denominator of the square root is
introduced because in the harmonic oscillator basis the matrix
elements of the $\sigma$-operators behave like $\sqrt{N-n_\pi}$, which
for large $N$ and small number of $\boldsymbol{\pi}$ bosons is
approximated by $\sqrt{N}$. {\it This approximate value of the
  $\boldsymbol{\sigma}$ operators is satisfied in any basis, with the
  condition that the average number of $\boldsymbol{\pi}$ bosons is
  much smaller than $N$} (though, the structure is particularly simple
in the harmonic oscillator basis).  Thus the $1/\sqrt{N}$ factor
cancels approximately the contributions due to the addition of the
$\boldsymbol{\sigma}^\dagger$ and $\boldsymbol{\sigma}$ operators. In
this form, the algebraic coordinate operator does not depend on the
basis used (the $\boldsymbol{r}^a_m$ can be applied to any kind of
basis) and nor on the Hamiltonian, thus, satisfying condition ii). For
very large $N$ the expressions of the physical and the algebraic
coordinate operators tend to each other, satisfying condition iii).

This definition agrees with \cite{lemus1,lemus2} where an algebraic
model for {\it atomic molecules} is discussed. Often (see for example
\cite{levit}) one defines the radial distance as a function of the
dynamical symmetry, relating it indirectly to the matrix element of
the dipole operator, without any further considerations. This violates
condition ii) above. We insist that the definition of the radial
coordinate operator has to be independent of the Hamiltonian in the
Hilbert space considered. The Hamiltonian determines if there is a
dynamical symmetry or not, which should be independent of the radial
coordinate operator, while the Hilbert space as such is independent of
the basis used.

The expectation value of the algebraic coordinate operator is
\begin{eqnarray}
\langle \boldsymbol{r}_m^a \rangle & = & \sqrt{ \frac{2N\hbar}{m\omega}}
\frac{\alpha_m}{\left(1+\alpha^2\right)} ~=~ r_m^a
~~~.
\label{ra}
\end{eqnarray}
We define this as the algebraic distance $r_m^a$, which is, by
definition, of the order of one.  Inverting this relation gives
\begin{eqnarray}
\frac{\alpha_m}{\left(1+\alpha^2\right)} & = &
\sqrt{ \frac{m\omega}{2N\hbar}} r_m^a
~~~,
\end{eqnarray}
which again provides the dependence of $\alpha_m$ on $N$. It suggests
we redefine $\alpha_m$ in terms of $\delta_m$ and $N$ as given above,
i.e.,
\begin{eqnarray}
\alpha_m & = & \frac{\delta_m}{\sqrt{N}} ~~~,
\end{eqnarray}
with $\delta_m$ given by
\begin{eqnarray}
\frac{\delta_m}{( 1+\frac{\delta^2}{N})} & = &
\sqrt{ \frac{m\omega}{2\hbar}} r_m^a
\label{x29}
~~~,
\end{eqnarray}
a dimensionless measure of the distance between the two nuclear
clusters.  We claim that this is a consistent way to define the
distance between clusters trough the variable $\delta_m$.

The validity of (\ref{ra}), the definition of $r_m^a$, depends on the
fluctuations of the related expectation value. This will be discussed
shortly.  The inter-cluster distance vector can always be chosen along
the z-axis.  The square of the variation ($\langle \boldsymbol{r}_0^{a
  2} \rangle - \langle \boldsymbol{r}_0^a \rangle^2$) can then also be
calculated, giving
\begin{eqnarray}
\langle (\boldsymbol{r}_0^{a})^{2} \rangle 
- \langle \boldsymbol{r}_0^a \rangle^2 & = &
-\left( \frac{4\hbar}{m\omega}\right) 
\frac{\left(\frac{\delta_0^2}{N}\right)}{\left(1+\frac{\delta^2}{N} \right)^2}
+ \left( \frac{\hbar}{2m\omega}\right)
\frac{1+{\textstyle\frac{\delta_0}{N}}}{1+{\textstyle\frac{\delta^2}{N}}}
\nonumber \\
& \rightarrow & \left( \frac{\hbar}{2m\omega}\right)
~~~,
\label{root}
\end{eqnarray}
where the arrow gives the limit for large $N$.  As long as the
expectation value of the algebraic distance operator is greater than
the square root of this expression, it is safe to identify the $r_0^a$
as the distance between the two clusters. The square root of
(\ref{root}) gives numbers of the order of 1 fm.

When applying the PACM the usage of another variable  
\begin{eqnarray}
\beta^2 & = & \frac{\alpha^2}{(1+\alpha^2)}
~~~.
\label{29}
\end{eqnarray}
will be found more favorable.  The range of $\beta^2$ is\linebreak $0\leq
\beta^2 \leq 1$, because the range of $\alpha$ is $0\leq \alpha^2 <
\infty$. The $\beta^2$ variable can be related to $\delta^2$ as it was
to $\alpha^2$.  The main reason to use $\beta$ is that within the PACM
it simplifies most expressions, as can be seen further below.  For the
SACM we will return, for convenience, to the variable $\alpha$.

Before closing this subsection we note that in the SACM the relation
between $\alpha$ and the relative distance of the two clusters is
given by $\alpha \sim \left(r-r_0\right)$, \cite{geom}, with $r_0 \sim
\sqrt{n_0}$. This means that there is a minimal difference in the
inter-cluster distance due to the presence of a minimal number of
$\pi$-bosons, i.e., clusters can not overlap completely. In the PACM,
$n_0=0$ and thus $r_0=0$.  Defining the minimum at $\alpha = 0$ as
the ``spherical'' minimum then loses its meaning, because in the SACM
it already corresponds to a minimal distance. Nevertheless, we will
continue to call a minimum at $\alpha = 0$ a ``spherical'' minimum and
$\alpha \neq 0$ as a ``deformed'' minimum.

\section{The geometrically mapped potential}
\label{four}

As discussed previously, the difference between the SACM and PACM 
model spaces 
manifests itself in the difference of the matrix 
elements of physical operators, even if the operators themselves 
are the same in the 
two approaches. In this section we determine the potential energy surfaces 
in both models and explore the relation between them.

\subsection{The SACM case: Pauli principle taken into account}
\label{four-one}

Applying the coherent state for the SACM to the Hamiltonian (\ref{H-tot}), 
one obtains the geometrically mapped potential 
\begin{eqnarray}
\langle \mbox{\boldmath$H$}\rangle &=&
{\cal C}(x,y)-(b+{\bar b})xy 
\left(
A(x,y)\alpha^2\frac{F_{11}\left( \alpha ^{2}\right) }{%
F_{00}\left( \alpha ^{2}\right) }
\right.
\nonumber \\
&&\left.
-B(x,y)\alpha^4\frac{%
F_{22}\left( \alpha ^{2}\right) }{F_{00}\left( \alpha ^{2}\right) } 
+\alpha^6\frac{F_{33}\left( \alpha ^{2}\right) }{%
F_{00}\left( \alpha ^{2}\right) }
\right.
\nonumber \\
&&\left.
-C(x,y)\alpha^2\frac{%
F_{20}^{N-2}\left( \alpha ^{2}\right) }{F_{00}\left( \alpha ^{2}\right) }%
\right)
\label{exp-h}
\end{eqnarray}
where 
\begin{align}
&{\cal C}(x,y) =
\nonumber \\
&\langle \left( a_{Clus}+a+bn_{0}\right) xy \mathit{C}
_{2}\left( \lambda _{C},\mu _{C}\right) +\gamma \boldsymbol{L_{C}}%
^{2}+\left( 1-xy\right) \boldsymbol{L_{C}}^{2}\rangle  
\nonumber \\
&+xyt\langle \boldsymbol{K}^{2}\rangle 
+\frac{c}{4}\left( N+n_{0}\right) \left( N+n_{0}-1\right) y\left(
1-x\right) 
\label{v-eff}
\end{align}
and the $F_{ij}(\alpha^2)$ functions are defined as 
\begin{align}
& F_{00}\left( \alpha ^{2}\right) =
\frac{\left( N!\right)^{2}}{\left( N+n_{0}\right) !}%
\nonumber \\
&
\times \sum_{k=n_{0}}^{N+n_{0}}\left( 
\begin{array}{c}
N+n_{0} \\ 
k%
\end{array}%
\right) \left[ \frac{k!}{(k-n_{0})!}\right] ^{2}\alpha^{2k}
& \nonumber \\
& F_{11}\left( \alpha ^{2}\right) =
\frac{\left( N!\right)^{2}}{\left( N+n_{0}-1\right) !}
 \nonumber \\
&
\times \sum_{k=m\alpha x(n_{0}-1,0)}^{N+n_{0}-1}\left( 
\begin{array}{c}
N+n_{0}-1 \\ 
k%
\end{array}%
\right) \left[ \frac{(k+1)!}{(k+1-n_{0})!}\right] ^{2}\alpha^{2k} 
& \nonumber \\
& F_{22}\left( \alpha ^{2}\right) = 
\frac{\left( N!\right)^{2}}{\left( N+n_{0}-2\right) !}
\nonumber \\
&
\times \sum_{k=m\alpha x(n_{0}-2,0)}^{N+n_{0}-2}\left( 
\begin{array}{c}
N+n_{0}-2 \\ 
k%
\end{array}%
\right) \left[ \frac{(k+2)!}{(k+2-n_{0})!}\right] ^{2}\alpha^{2k}
& \nonumber \\
& F_{20}^{N-2}\left( \alpha ^{2}\right) =
\frac{\left( N!\right)^{2}}{\left( N+n_{0}-2\right) !}%
\nonumber \\
&
\times \sum_{k=n_{0}}^{N+n_{0}-2}\left( 
\begin{array}{c}
N+n_{0}-2 \\ 
k%
\end{array}%
\right) \left[ \frac{k!}{(k-n_{0})!}\right]
\nonumber \\
&
\hspace{1.5cm}\times \left[ \frac{(k+2)!}{(k+2-n_{0})!}\right]\alpha^{2k}
& \nonumber \\
& F_{33}\left( \alpha ^{2}\right) =
\frac{\left( N!\right)^{2}}{\left( N+n_{0}-3\right) !}
\nonumber \\
&
\times \sum_{k=m\alpha x(n_{0}-3,0)}^{N+n_{0}-3}\left( 
\begin{array}{c}
N+n_{0}-3 \\ 
k%
\end{array}%
\right) \left[ \frac{(k+3)!}{(k+3-n_{0})!}\right] ^{2}\alpha^{2k}
~.
\end{align}
Further, the constants appearing in (\ref{exp-h}) are defined as 
\begin{eqnarray}
A(x,y) &=&-\frac{1}{\left( b+\overline{b}~\right) xy}
\Big( 
\hbar \omega \left[ yx+1-y\right] 
\nonumber\\
&&
+2\left[ \gamma +(1-yx)a_{R}^{(1)}\right]
\nonumber\\
&&
\left.
+xy\left[ a-b\right] \left[ 4+\Gamma _{1}+\Gamma _{2}\right] 
\right.
\nonumber\\
&&
\left.
+4xy\left[ \overline{a}-\overline{b}~\right]  
+xybn_{0}\left[ 4+\Gamma _{1}+\Gamma _{2}\right]
+4xyn_{0}\overline{b}
\right.
\nonumber\\
&&
\left.
-bxy\mathit{C}_{2}\left( \lambda _{C},\mu _{C}\right) -\frac{c}{2}y
\left(1-x\right) \left( N+n_{0}-1\right) 
\right) 
\nonumber \\
B(x,y) &=& \frac{1}{\left( b+\overline{b}~\right) xy}
\Big( xy\left[ a+\overline{a}-6b-6\overline{b}
\right.
\nonumber\\
&&
\left. \left.
-b\left\{ \Gamma_{1}+\Gamma _{2}\right\} +n_{0}\left( b+\overline{b}~\right)
\right] +\frac{c}{2}y
\left( 1-x\right) \right) 
\nonumber
\\
C(x,y) &=&-\frac{\frac{c}{2}y\left( 1-x\right) }{\left( b+\overline{b}~\right)
xy} \:\:\:,
\label{abc}
\end{eqnarray}
where $\Gamma_k$, according to \cite{geom}, is given by
\begin{eqnarray}
\Gamma_k &=&\langle (\lambda _{k},\mu_{k})|
\boldsymbol{Q}_{m}^{Cluster(k)}|(\lambda_{k},\mu_{k})\rangle
\nonumber \\
&=&
\sqrt{\frac{5}{\pi}}\left[n_{k}+\frac{3}{2}\left(A_{k}-1\right)\right]
\alpha_{2m}\left(k\right) \nonumber \\
& = & \sqrt{\frac{5}{\pi}}N_{0,k}\beta_k
~~~.
\label{gamma1}
\end{eqnarray}
This was obtained by a geometric mapping of the symplectic model
\cite{sympl, troltenier,appl-map}).  The $(\lambda_k,\mu_k)$ denotes
the $SU(3)$ irrep of the deformed cluster number $k$.  In the case
that it is spherical, $\Gamma_k=0$. The $N_{0k}$ is the sum of the
total number of quanta $n_k$ of the deformed cluster plus
$\frac{3}{2}(A_k-1)$, where $A_k$ is the number of nucleons in the
$k^{\text{th}}$ cluster. This last term is the zero-point energy with the
contribution of the center of mass already extracted.  The
$\alpha_{2m}(k)$ is the deformation variable of cluster number $k$. In
(\ref{gamma1}) we used only the $m=0$ component of $\alpha_{2m}(k)$
and defined it $\beta_k$, the deformation of cluster $k$ 
(not to be confused with the $\beta$ variable appearing in 
Eq.~(\ref{29})). 
This implies
that the deformed cluster is assumed to be axially symmetric and it is
in line with the inter-cluster z-axis, which connects both
clusters. When the z-axis of the deformed cluster is inclined with
respect to the molecular z-axis by an angle $\theta$, the deformation
value $\beta_k$ is multiplied by a matrix element of the rotation
matrix, which only changes the numerical value of $\beta_k$, i.e.,
$\beta_k^\prime = d_{00}^2(\theta)\beta_k$.  For simplicity we do not
include these orientations in the discussion.  Furthermore, it will
not change the basic results.

In discussing the phase transitions it is possible to choose as the
independent parameters of the theory $A$, $B$ and $C$, which
themselves are the functions of all interaction parameters of the
theory. This structure will be used in the second paper, investigating
the possible phase transitions and the phase diagram.

For $x=0$ in (\ref{H-tot}), i.e. in the case of the $SO(4)$ to $SO(3)$ 
phase transition, the discussion has to be modified due to the $xy$ factor 
appearing in the denominators in (\ref{abc}). 
In this case the potential maps to
\begin{eqnarray}
V &=&\langle \boldsymbol{H}\rangle 
\nonumber \\
&\rightarrow& \left( \overline{A}%
\alpha ^{2}\frac{F_{11}\left( \alpha ^{2}\right) }{F_{00}\left( \alpha
^{2}\right) }-\overline{B}\alpha ^{4}\frac{F_{22}\left( \alpha ^{2}\right) }{%
F_{00}\left( \alpha ^{2}\right) }-\overline{C}\alpha ^{4}\frac{%
F_{20}^{N-2}\left( \alpha ^{2}\right) }{F_{00}\left( \alpha ^{2}\right) }%
\right) 
\nonumber \\
&&+ {\cal C} 
\end{eqnarray}
with
\begin{eqnarray}
\overline{A} &=&\left( \hbar \omega \left[ 1-y\right] +2\left[ \gamma
+a_{R}^{(1)}\right] -\frac{c}{2}y\left( N+n_{0}-1\right) \right) \nonumber \\ \\
\overline{B} &=&-\left( \frac{c}{2}y\right)  \\
\overline{C} &=&\frac{c}{2}y=-\overline{B}
~~~.
\end{eqnarray}
In this case only two independent parameters $\overline{A}$ and 
$\overline{B}$ appear.

Note that for $c>0$, the $C$ is positive (remember that $-b>0$). When
$C<0$, the situation corresponds in the $SO(4)$ limit to a ground
state where all bosons are decoupled and the highest state is the one
where all bosons are coupled in pairs.

In (\ref{exp-h}) we have to add a constant term, such that the
geometrically mapped potential is zero at $\alpha=0$. This is a
permitted renormalization of the zero-point energy. This constant will
be determined further below.

It also has to be noted that the $SO(4)$ dynamical symmetry needs
special care due to the truncation of the $SU(3)$ basis required by
the Pauli principle.  Eliminating these components leads to the
destruction of the $SO(4)$ dynamical symmetry. However, we will still
denote it a $SO(4)$ dynamical symmetry, because the operators in the
Hamiltonian will be the same. Instead of the $SO(4)$ basis we will
stay within the $SU(3)$ basis, because only there can the Pauli
principle be implemented easily, canceling all states with
$n_\pi<n_0$. Note also that the total number of bosons is now
$(N+n_0)$ and {\it not} just $N$.

A very useful consideration is the investigation of the potential in
the $\alpha\rightarrow\infty$ and $\alpha \rightarrow 0$ limits.  In
the first limit we will see that the potential approaches a constant
value depending on $(N+n_0)$, which is due to the finite size of the
boson space. For large values of $\alpha$ the coherent state contains
only $\pi$-bosons and cannot increase the energy any further.  The
second limit ($\alpha \rightarrow 0$) is necessary to adjust
$V(\alpha=0)=0$.  These consideration will be important in the second
paper, when the general structure of the SACM phase diagram will be
discussed.

\noindent
{\bf i) Limit} $\alpha\rightarrow\infty$: 

The relevant formulas are
\begin{eqnarray} 
\alpha^2 \frac{F_{11}}{F_{00}} & \rightarrow &
\left(N+n_0\right)
\nonumber \\
\alpha^4 \frac{F_{22}}{F_{00}} & \rightarrow &
\left(N+n_0\right)\left(N+n_0-1\right)
\nonumber \\
\alpha^6 \frac{F_{33}}{F_{00}} & \rightarrow &
\left(N+n_0\right)\left(N+n_0-1\right)\left(N+n_0-2\right)
\nonumber \\
\alpha^2 \frac{F_{20}^{N-2}}{F_{00}} & \rightarrow &
N\left(N-1\right)
\frac{1}{\alpha^2} ~\rightarrow~0 ~~~ . 
\label{alpha-ifnty-1}
\end{eqnarray}
With this, the limit of the complete geometric potential
(\ref{v-eff}) is given by
\begin{eqnarray}
V & \rightarrow &
{\cal C} - \left( b + {\bar b} \right) xy \left\{ A\left(N+n_0\right)
-B\left(N+n_0\right)\left(N+n_0-1\right)
\right.
\nonumber \\
&&\left. +\left(N+n_0\right)\left(N+n_0-1\right)\left(N+n_0-2\right)
\right\}
~~~.
\end{eqnarray}
Depending on the signs and values of $A$ and $B$, this limit is either
positive or negative. For the positive value the limit for
$(N+n_0)\rightarrow\infty$ is then $+\infty$, leading to a stable
potential, while if it is negative the limit leads to $-\infty$,
leading to an unstable potential.

\noindent
{\bf ii) Limit} $\alpha \rightarrow 0$: 

The relevant formulas are
\begin{eqnarray}
\alpha^2 \frac{F_{11}}{F_{00}} & \rightarrow & n_0
\nonumber \\
\alpha^4 \frac{F_{22}}{F_{00}} & \rightarrow & n_0\left( n_0-1\right)
\nonumber \\
\alpha^6 \frac{F_{33}}{F_{00}} & \rightarrow & n_0\left( n_0-1\right)
\left( n_0-2\right)
\nonumber \\
\alpha^2 \frac{F_{20}^{N-2}}{F_{00}} & \rightarrow & N\left(N-1\right)
(n_0+1)(n_0+2)
\frac{\alpha^2}{2} ~\rightarrow~0
~~~.
\nonumber \\
\label{alpha-inty-2}
\end{eqnarray}
With this, the limit of the complete geometric potential
(\ref{v-eff}) is given by
\begin{eqnarray}
V(\alpha = 0) & \rightarrow & {\cal C} 
-\left( b + {\bar b} \right) xy\left\{ An_0 - B n_0\left( n_0-1\right)
\right.
\nonumber \\
&&\hspace{1.5cm}
\left.
+ n_0\left( n_0-1\right)\left( n_0-2\right)
\right\}
~~~,
\end{eqnarray}
which is independent of $N$. This result can be used to adjust the 
potential to zero at $\alpha=0$.

\subsection{The PACM case: Pauli principle not taken into account}
\label{four-two}

Using the Hamiltonian as introduced in the section \ref{two} and the coherent
state of section \ref{three}, for the case when the Pauli exclusion principle 
is not taken into account, the potential is obtained by calculating
the expectation value of the Hamiltonian as
\begin{align}
&\langle \boldsymbol{H}\rangle = V(\beta)
\nonumber \\
&= \left( a_{Clus}+a\right) xy {\cal C}_{2}\left( \lambda _{C},\mu _{C}\right)
+\frac{c}{4}N\left( N-1\right) y\left( 1-x\right) 
\nonumber \\
&~~~
+N\beta^2 \Big[ \left(\hbar\omega\left( xy+1-y\right)
+ \left( a - b\right)\left(4+\Gamma_{1} + \Gamma_{2} \right)
\right. 
\nonumber \\
&~~~~~~~
\left. 
+4\left( {\bar a} - {\bar b}\right)
-b\,{\cal C}_{2} \left(\lambda_{C} , \mu_{C} \right)  \right)
 - \left( 1-x \right) y c\left( N-1 \right)
\nonumber \\
&~~~~~~~
\left. 
+2\left( \gamma +(1-xy)a_{R}^{(1)}\right)
\right]
\nonumber \\
&~~~
+N \left( N-1 \right) \beta^{4} \left[ xy
\left( a+{\bar a}-6b-6{\bar b}
-b\left( \Gamma_{1} + \Gamma_{2}\right)  \right) 
\right.
\nonumber \\
&~~~~~~~
\left.
+ \left( 1-x \right) y c \right]
\nonumber \\
&~~~
- N \left( N-1 \right)  \left( N-2 \right) \beta
^{6}xy\left(b+{\bar b}\right)+{\cal C}_{2} \left (\lambda_{C} , 
\mu_{C} \right) a xy
\nonumber \\
&~~~~~~~
+ \frac{1}{4}\, \left( 1-x \right) ycN \left( N-1 \right)
~~~.
\label{h-v}
\end{align}

Defining
\begin{eqnarray}
A &=& -\left[\left( b+\overline{b}~\right) xy(N-1)(N-2)\right]^{-1}
\nonumber \\
&&\times 
\left[ 
\hbar \omega \left( yx+1-y \right) +2\left( \gamma +(1-yx)a_{R}^{(1)}\right)
\right.
\nonumber \\
&& 
~~~+4xy\left( \overline{a}-\overline{b}\,\right) 
 +xy\left( a-b\right) \left( 4+\Gamma _{1}+\Gamma _{2}\right) 
\nonumber \\
&&
~~~-bxy\mathit{C}_{2}\left( \lambda _{C},\mu _{C}\right) -y\left( 1-x\right)
c\left( N-1\right) 
\Big]  
\nonumber \\
B &=&
\frac{ xy\left( a+\overline{a}-6\left( b+\overline{b}\,\right)
-b\left( \Gamma _{1}+\Gamma _{2}\right) \right) +cy\left( 1-x\right) }
{(N-2) \left( b+\overline{b}~\right) xy }
\nonumber \\
{\cal C}&=&\left< \left( a_{Clus}+a\right)
xy \mathit{C}_{2}\left(\lambda _{C},\mu _{C}\right) 
+\gamma \boldsymbol{L_{C}}^{2}
+\left(1-xy\right)
 \boldsymbol{L_{C}}^{2}\right> 
\nonumber \\
&&
 +xyt\langle \boldsymbol{K}^{2}\rangle 
+\frac{c}{4}N\left( N-1\right) y\left( 1-x\right) \:\:\:,
\label{v1}
\end{eqnarray}
the potential acquires the form \cite{lorena-T,cocoyoc-2011}
\begin{eqnarray}
V &=& N(N-1)(N-2) (-(b+{\bar b})xy) \left\{  A\beta^2 -B \beta^4 
+ \beta^6 \right\}
\nonumber \\
&& + {\cal C}
~~~,
\end{eqnarray}
which allows us to define a new, normalized potential
\begin{eqnarray}
{\widetilde V} & = & \left\{ A\beta^2 -B \beta^4 + \beta^6 \right\}
~~~.
\label{vtilde2}
\end{eqnarray}
In the definition of ${\widetilde V}$ we extracted the factor
$(b+{\bar b})xy$, such that there appears no factor in front of the
$\beta^6$ term. This poses no problem as long as $x$ is varied from 0
to 1.  In the limit of $x \rightarrow 0$ the $A$ and $B$ values
also approach $\pm\infty$, depending on the sign.  However, for the
$SO(4)$ to $SO(3)$ transition, the $x$ value is always zero.  For this
case we include the $x$ value within the parenthesis, yielding a
vanishing factor of the sextic term.

Comparing the potentials obtained from the same Hamiltonian in the
SACM and PACM approaches leads to a remarkable finding. The potential
in the SACM framework is rather different from its PACM counterpart,
however, a similar potential can also be generated within the latter
framework too. This can be achieved by including higher-order
interactions of the type
$F_1(\boldsymbol{n}_\pi)/F_2(\boldsymbol{n}_\pi)$, with appropriate
functions $F_k(\boldsymbol{n}_\pi )$. This demonstrates that observing
the Pauli exclusion principle acts {\it as if} one used high-order
interactions in a model which does not observe the Pauli exclusion
principle. In fact, the non-linear terms simulate the presence of the
Pauli exclusion principle.

\section{Conclusions}
\label{five}

In order to investigate possible phase transitions between different
limits corresponding to various dynamical symmetries, we reparametrized
the Hamiltonian of the Semimicroscopic Algebraic Cluster Model (SACM)
such that it allowed interpolation between the three possible limits.
These were the {\it strong coupling limit} ($SU(3)$), the {\it
  deformed limit} ($SO(4)$) and the {\it weak coupling limit}
($SO(3)$).  The latter limit was proposed in the present work and it
differs from the strong coupling limit in the level on which the
interaction terms of the relative motion and those of the internal
cluster structure are coupled: in the weak coupling limit this is done
on the $SO(3)$ (i.e. angular momentum) level, while in the strong
coupling limit the $SU(3)$ algebra plays a role, introducing,
e.g. quadrupole--quadrupole interaction between the two sectors. In
the case of a system with two spherical clusters the weak coupling
Hamiltonian is a simplified version of the strong coupling one, so it
does not stand as a separate limit in itself.  The $SO(4)$ limit also
has its limitations due to the truncation of the model space in the
$n_{\pi}$ quantum number.

The Phenomenological Algebraic Cluster Model (PACM) was introduced as a
special limit of the SACM with the minimal number of the $\pi$ bosons
set to zero. This choice corresponds to neglecting the effects of the
Pauli exclusion principle. Although this means giving up a fundamental
physical requirement, this decision was inspired by the fact that the
formalism of the PACM is closer to other similar models using the
coherent state method. It appears instructive to study the differences
and similarities between the SACM and PACM within this latter
approach.

The present work is meant to be the basis for a further study in which
phase transitions are investigated by interpolating between two
dynamical symmetry limits.  This method requires the application of
large boson numbers, so as another new ingredient, the Hamiltonian was
implemented with a third-order term in order to stabilize the energy
spectrum in this situation.  The potential energy surface was
constructed in terms of a variable controlling the relative distance
of the clusters. This was done both in the SACM and PACM framework. It
was found that the potential obtained from the SACM can be reproduced
within the PACM approach too by including higher-order terms in the
Hamiltonian. This indicates that studying only the Hamiltonian, the
effects of the Pauli principle can be simulated by higher-order
interactions.

The present results will be used in a forthcoming publication that
focuses on phase transitions beween phases determined by different
dynamical symmetries of the SACM and the PACM.

\section*{APPENDIX A: The Coherent State for the SACM}

We choose the most general structure for the coherent state, allowing
arbitrary parameters, $a_m$, which only coincide with $\alpha_m$ when
the static problem is considered.  This will be important
in future work, when we
intend to treat the cranking formalism within the PACM and SACM,
similar to the formalism presented in \cite{schaaser1,schaaser2}.
Nevertheless, as long as we are only interested in the potential
energy surface for systems without rotation, the parameters $\alpha_m$
will form a simple tensor.

We use the definition
\begin{eqnarray}
\left(\mbox{\boldmath$\alpha$}\cdot\mbox{\boldmath$\pi$}^{\dagger}\right)
& = & \sum_{m}\alpha_{m}\mbox{\boldmath$\pi$}_{m}^{\dagger}
~~~.
\end{eqnarray}
The $\alpha_{m}$ are in general complex and arbitrary. The complex
conjugate is denoted by $\alpha_{m}^{*}$. We also use
\begin{eqnarray}
\tilde{\alpha}_{m} & = & (-1)^{1-m}\alpha_{-m}~~~.
\end{eqnarray}
This will be important when we apply
$\boldsymbol{\pi}^{m}~=~(-1)^{1-m}\boldsymbol{\pi}_{-m}$ to the
coherent state on the right.

The conjugate coherent state is given by
\begin{eqnarray}
\langle \boldsymbol{\alpha}| & = & \mathcal{N}_{Nn_{0}}\langle0|\left[
\boldsymbol{\sigma}+\left(\mbox{\boldmath$\alpha$}^{*}\cdot\mbox{\boldmath$\pi$}
\right)\right]^{N}\left(\mbox{\boldmath$\alpha$}^{*}\cdot\mbox{\boldmath$\pi$}
\right)^{n_{0}}~~~,
\end{eqnarray}
with
\begin{eqnarray}
\left(\mbox{\boldmath$\alpha$}^{*}\cdot\mbox{\boldmath$\pi$}\right) & = &
\sum_{m}\alpha_{m}^{*}\mbox{\boldmath$\pi$}^{m}~~~.
\end{eqnarray}
Thus, the $\boldsymbol{\pi}_{m}^{\dagger}$ acts on the left
as an annihilation operator.
Note that here we do not assume a tensorial behavior of the
$\alpha_m$, contary to what we used in the body of the paper.
In order to relate
this $\alpha_m$ to the one used in the paper, we have to assume
$\alpha_m^*$ = $\left(-1\right)^m \alpha_{-m}$. This is justified for
a static problem, as discussed in the paper. The situation changes,
when for example the cranking formalism is applied or not only the
potential is intented to derive but also the kinetic energy.

$\mathcal{N}_{Nn_{0}}$ is the normalization factor, given by
\begin{equation}
\mathcal{N}_{Nn_{0}}^{-2} =
\frac{N!^{2}}{(N+n_{0})!}\frac{{\rm d}^{n_{0}}}{{\rm d}\gamma_{1}^{n_{0}}}\
\frac{{\rm d}^{n_{0}}}{{\rm d}\gamma_{2}^{n_{0}}}
\left[1+\gamma_{1}\gamma_{2}(\boldsymbol{\alpha}^{*}\cdot
 \boldsymbol{\alpha})\right]^{N+n_{0}}
~,
\end{equation}
taken at $\gamma_{1}=\gamma_{2}=1$, after performing the derivation.

Acting with $\mbox{\boldmath$\pi$}_{m}^{\dagger}$ to the left,
commutators of the type
$\left[\boldsymbol{\pi}^{m_{1}},\boldsymbol{\pi}_{m_{2}}^{\dagger}\right]$
appear and will give expressions proportional to $\alpha_{m}^{*}$.
However, acting with $\mbox{\boldmath$\pi$}_{m}$, as it appears in a
coupled expression, to the right, it will give expressions
proportional to $\tilde{\alpha}_{m}=(-1)^{1-m}\alpha_{-m}$, because we
have first to lift the index of the annihilation operator, obtaining
$(-1)^{1-m}\mbox{\boldmath$\pi$}^{-m}$.  Note that
\begin{eqnarray}
\left(\boldsymbol{\alpha}^{*}\cdot \boldsymbol{\alpha}\right) & = &
\sum_{m}\alpha_{m}^{*}\alpha_{m}~=~\sum_{m}|\alpha_{m}|^{2}
\end{eqnarray}
is a real number.

The formulas are now similar to those of Ref. \cite{geom}, but without
the use of a possible tensor character of the $\alpha_{m}$ and with
the appearance of complex conjugate $\alpha_{m}^{*}$ and
$\tilde{\alpha}_{m}$.  One of the interesting matrix element is given
by \cite{geom}
\begin{multline}
\left< \left[\mbox{\boldmath$\pi$}^{\dagger}\otimes\mbox{\boldmath$\pi$}
\right]_{m}^{\left[S\right]}\right> =
(N+n_{0})\left[\mbox{\boldmath$\alpha$}^{*}\times
\tilde{\mbox{\boldmath$\alpha$}}\right]_{m}^{S}\mathcal{N}_{Nn_{0}}^{2}
\frac{N!^{2}}{(N+n_{0})!}\nonumber \\
\frac{{\rm d}^{n_{0}}}{{\rm d}\gamma_{1}^{n_{0}}}\frac{{\rm d}^{n_{0}}}
{{\rm d}\gamma_{2}^{n_{0}}}\gamma_{1}\gamma_{2}
\left[1+\gamma_{1}\gamma_{2}(\boldsymbol{\alpha}^{*}\cdot
 \boldsymbol{\alpha})\right]^{N+n_{0}-1}
~.
\end{multline}
We have
\begin{eqnarray}
\left[\mbox{\boldmath$\alpha$}^{*}\times\tilde{\mbox{\boldmath$\alpha$}}
\right]_{m}^{S} & = &
\sum_{m_{1}m_{2}}(1m_{1},1m_{2}|S m)
\alpha_{m_{1}}^{*}\tilde{\alpha}_{m_{2}}~~~,
\end{eqnarray}
where the coupling sign ``$\times$" instead of ``$\otimes$" was used
in order to indicate that we do not couple tensors. This is just a
short-hand notation.

In the geometrical mapping one has to take into account that
\begin{eqnarray}
(\mbox{\boldmath$\alpha$}\cdot\mbox{\boldmath$\pi$}^{\dagger}) & = &
\alpha_{0}\mbox{\boldmath$\pi$}_{0}^{\dagger}
 +\alpha_{1}\mbox{\boldmath$\pi$}_{+1}^{\dagger}
+\alpha_{-1}\mbox{\boldmath$\pi$}_{-1}^{\dagger}
\end{eqnarray}
and thus
\begin{eqnarray}
\left[\mbox{\boldmath$\pi$}_{m},
(\mbox{\boldmath$\alpha$}\cdot\mbox{\boldmath$\pi$}^{\dagger})\right] & = &
(-1)^{1-m}\alpha_{-m}~=~\tilde{\alpha}_{m}\nonumber \\
\left[(\mbox{\boldmath$\alpha$}^{*}\cdot\mbox{\boldmath$\pi$}),
\mbox{\boldmath$\pi$}_{m}^{\dagger}\right] & = &
\alpha_{m}^{*}
~~~.
\end{eqnarray}
Note the difference in the phase.

In this sense, the mapping of different operators is completely
parallel to the one given in \cite{geom}, with the exception of the
definition in the coupling of $\alpha_{m}$. The geometrical mapping of
more relevant operators is given by
\begin{align}
& \langle\mbox{\boldmath$\alpha$}|\mbox{\boldmath$\sigma$}^{\dagger}
 \mbox{\boldmath$\pi$}_{m}|\mbox{\boldmath$\alpha$}\rangle=
 \nonumber \\
& (-1)^{1-m}(N+n_{0})\tilde{\alpha}_{m}\mathcal{N}_{Nn_{0}}^{2}
 \frac{(N!)^{2}}{(N+n_{0})!}
\nonumber \\
& \hspace{1.5cm} \times \frac{{\rm d}^{n_{0}}}{{\rm d}\gamma_{1}^{n_{0}}}
\frac{{\rm d}^{n_{0}}}{{\rm d}\gamma_{2}^{n_{0}}}
 \left[ \gamma_{2}\left[1+\gamma_{1}\gamma_{2}
 (\mbox{\boldmath$\alpha$}^{\ast}\cdot\mbox{\boldmath$\alpha$})\right]^{N+n_{0}-1}
 \right]
 \nonumber \\
 & \langle\mbox{\boldmath$\alpha$}|\mbox{\boldmath$\pi$}_{m}^{\dagger}
 \mbox{\boldmath$\sigma$}|\mbox{\boldmath$\alpha$}\rangle=
 \nonumber \\
 & (-1)^{1-m}(N+n_{0})\alpha_{m}^{\ast}\mathcal{N}_{Nn_{0}}^{2}
 \frac{(N!)^{2}}{(N+n_{0})!}
\nonumber \\
& \hspace{1.5cm} \times
 \frac{{\rm d}^{n_{0}}}{{\rm d}\gamma_{1}^{n_{0}}}\frac{{\rm d}^{n_{0}}}
 {{\rm d}\gamma_{2}^{n_{0}}}\left[\gamma_{2}
 \left[1+\gamma_{1}\gamma_{2}
 (\mbox{\boldmath$\alpha$}^{\ast}\cdot\mbox{\boldmath$\alpha$})\right]^{N+n_{0}-1}
 \right]
 \nonumber \\
 & \langle\mbox{\boldmath$\alpha$}|\mbox{\boldmath$\sigma$}^{\dagger}
 \mbox{\boldmath$\sigma$}|\mbox{\boldmath$\alpha$}\rangle=
 N^{2}\frac{\mathcal{N}_{Nn_{0}}^{2}}{\mathcal{N}_{(N-1)n_{0}}^{2}}
 \nonumber \\
 & \langle\mbox{\boldmath$\alpha$}
 |\left[\left[\mbox{\boldmath$\pi$}^{\dagger}
 \otimes\mbox{\boldmath$\pi$}^{\dagger}\right]^{\left[S_{1}\right]}
 \otimes\left[\mbox{\boldmath$\pi$}
 \otimes\mbox{\boldmath$\pi$}\right]^{\left[S_{2}\right]}
 \right]_{m}^{\left[S_{3}\right]}|\mbox{\boldmath$\alpha$}\rangle=
 \nonumber \\
 & (N+n_{0})(N+n_{0}-1)\mathcal{N}_{Nn_{0}}^{2}
\frac{(N!)^{2}}{(N+n_{0})!}
 \nonumber \\
&\hspace{1.5cm}\times 
\left[\left[\mbox{\boldmath$\alpha$}^{\ast}\times\mbox{\boldmath$\alpha$}^{\ast}
 \right]^{\left[S_{1}\right]}\times\left[\tilde{\mbox{\boldmath$\alpha$}}
 \times\tilde{\mbox{\boldmath$\alpha$}}\right]^{\left[S_{2}\right]}
 \right]_{m}^{\left[S_{3}\right]}
\nonumber \\
& \hspace{1.5cm} \times
\frac{{\rm d}^{n_{0}}}
 {{\rm d}\gamma_{1}^{n_{0}}}\frac{{\rm d}^{n_{0}}}{{\rm d}\gamma_{2}^{n_{0}}}
 (\gamma_{1}\gamma_{2})^{2}
 \left[1+\gamma_{1}\gamma_{2}(\alpha^{\ast}\cdot \alpha)\right]^{N+n_{0}-2}
 \nonumber \\
 & \langle\mbox{\boldmath$\alpha$}|\left[\mbox{\boldmath$\pi$}^{\dagger}
 \otimes\mbox{\boldmath$\pi$}^{\dagger}\right]_{\mu}^{\left[S\right]}
 (\mbox{\boldmath$\sigma$})^{2}
 |\mbox{\boldmath$\alpha$}\rangle=
 \nonumber \\
 & (N+n_{0})(N+n_{0}-1)\mathcal{N}_{Nn_{0}}^{2}
 \frac{(N!)^{2}}{(N+n_{0})!}
 \left[\mbox{\boldmath$\alpha$}^{\ast}\times\mbox{\boldmath$\alpha$}^{\ast}
 \right]_{\mu}^{\left[S\right]}
\nonumber \\
& \hspace{1.5cm} \times
\frac{{\rm d}^{n_{0}}}{{\rm d}\gamma_{1}^{n_{0}}}
\frac{{\rm d}^{n_{0}}}
 {{\rm d}\gamma_{2}^{n_{0}}}\left\{ \gamma_{2}^{2}
 \left[1+\gamma_{1}\gamma_{2}(\alpha^{\ast}\cdot \alpha)\right]^{N+n_{0}-2}
\right\}
 \nonumber \\
 & \langle\mbox{\boldmath$\alpha$}|(\mbox{\boldmath$\sigma$}^{\dagger})^{2}
 \left[\mbox{\boldmath$\pi$}\otimes\mbox{\boldmath$\pi$}
 \right]_{\mu}^{\left[S\right]}|\mbox{\boldmath$\alpha$}\rangle=
 \nonumber \\
 & (N+n_{0})(N+n_{0}-1)\mathcal{N}_{Nn_{0}}^{2}
 \frac{(N!)^{2}}{(N+n_{0})!}
 \left[\tilde{\mbox{\boldmath$\alpha$}}\times
 \tilde{\mbox{\boldmath$\alpha$}}\right]_{\mu}^{\left[S\right]}
\nonumber \\
& \hspace{1.5cm} \times 
\frac{{\rm d}^{n_{0}}}{{\rm d}\gamma_{1}^{n_{0}}}\frac{{\rm d}^{n_{0}}}
 {{\rm d}\gamma_{2}^{n_{0}}}
 \left\{ \gamma_{1}^{2}\left[1
 +\gamma_{1}\gamma_{2}(\alpha^{\ast}\cdot \alpha)\right]^{N+n_{0}-2}\right\}
 \nonumber \\
 & \langle\mbox{\boldmath$\alpha$}|(\boldsymbol{\sigma}^{\dagger})^{2}
 (\boldsymbol{\sigma})^{2}|\mbox{\boldmath$\alpha$}\rangle=
 N(N-1)\frac{\mathcal{N}_{Nn_{0}}^{2}}{\mathcal{N}_{(N-2)n_{0}}^{2}} &
 ~~~.
\end{align}
These equations also give us the mapping of $\boldsymbol{n}_\pi$ and
$\boldsymbol{n}_\pi^2$ as special cases. For completeness we also give
the mapping of $\boldsymbol{n}_\pi^3$, which is
\begin{align}
&\left< {\boldsymbol{\alpha}} \right| \boldsymbol{n}_{\pi}^{2}
+2\sum_{i,j}\boldsymbol{\pi}_{i}^{\dagger}\boldsymbol{\pi}_{j}^{\dagger}
\boldsymbol{\pi}^{i}\boldsymbol{\pi}^{j}
+ \sum_{i,j,k}\boldsymbol{\pi}_{i}^{\dagger}
\boldsymbol{\pi}_{j}^{\dagger}\boldsymbol{\pi}_{k}^{\dagger}
\boldsymbol{\pi}^{i}\boldsymbol{\pi}^{j}\boldsymbol{\pi}^{k} 
\left| {\boldsymbol{\alpha}} \right>
\nonumber \\
& = \left< {\boldsymbol{\alpha}} \right|
3\left\{ \left[{\ \mbox{\boldmath$\pi$}}^{\dagger}\otimes{\ %
\mbox{\boldmath$\pi$}}^{\dagger}\right]_{0}^{0}\left[{\ \mbox{\boldmath$\pi$}%
}\otimes{\ \mbox{\boldmath$\pi$}}\right]_{0}^{0} 
\right.
\nonumber \\
&\left. \hspace{1.8cm}
+\, \sqrt{5}\left[\left[{\ %
\mbox{\boldmath$\pi$}}^{\dagger}\otimes{\ \mbox{\boldmath$\pi$}}^{\dagger}%
\right]^{2}\otimes\left[{\ \mbox{\boldmath$\pi$}}\otimes{\ %
\mbox{\boldmath$\pi$}}\right]^{2}\right]_{0}^{0}\right\}
\left| {\boldsymbol{\alpha}} \right> 
\nonumber \\
& + \left< {\boldsymbol{\alpha}} \right|
\sum_{i,j,k}\boldsymbol{\pi}_{i}^{\dagger}\boldsymbol{\pi}_{j}^{\dagger}
\boldsymbol{\pi}_{k}^{\dagger}\boldsymbol{\pi}^{i}\boldsymbol{\pi}^{j}
\boldsymbol{\pi}^{k}
+
\boldsymbol{n}_{\pi} \left| {\boldsymbol{\alpha}} \right> ~~~.
\nonumber \\
\end{align}

Note that the mapping is more complicated than when no Pauli exclusion 
principle is taken into account ($n_0=0$),
due to the distinct property of $\alpha_{m}$. (It is not a tensor
anymore.) Also note that
\begin{eqnarray}
\left[\tilde{\mbox{\boldmath$\alpha$}}\times
\tilde{\mbox{\boldmath$\alpha$}}\right]_{0}^{\left[0\right]} & = &
\frac{1}{\sqrt{3}}\sum_{m}(-1)^{1-m}\alpha_{m}\alpha_{-m}
\nonumber \\
\left[\mbox{\boldmath$\alpha$}^{*}\times
\mbox{\boldmath$\alpha$}^{*}\right]_{0}^{\left[0\right]} & = &
\frac{1}{\sqrt{3}}\sum_{m}(-1)^{1-m}\alpha_{m}^{*}\alpha_{-m}^{*}
~~~.
\end{eqnarray}
Thus, the sum of both is real. Because they always appear in a sum in
the expectation value of the Hamiltonian with respect to the coherent
state, the expectation value is always real. 
This is a remarkable sign of consistency. 

The mapping, concerning the individual clusters, is the same as given
in \cite{geom}. There, one has to take into account that the coherent
state acquires the form of a direct product of the state describing
the relative motion and the one giving the cluster coupling
\begin{eqnarray}
|\mbox{\boldmath$\alpha$}\rangle\left|
\left[C_{1}\times C_{2}\right]^{C}\right\rangle~~~,
\end{eqnarray}
where the last factor refers to the coupling of the two cluster
states, \textit{which is fixed} \cite{geom}.

Next we have to expand the above expressions in powers of
$(\mbox{\boldmath$\alpha$}\cdot\mbox{\boldmath$\alpha$})$.
In \cite{geom} the $n_{0}$ was neglected compared to $N$.
Here, we will take into account the contributions of $n_{0}$.
The list of expansions is
\begin{align}
 & \frac{{\rm d}^{n_{0}}}{{\rm d}\gamma_{1}^{n_{0}}}
 \frac{{\rm d}^{n_{0}}}{{\rm d}\gamma_{2}^{n_{0}}}
 \left[1+\gamma_{1}\gamma_{2}(\boldsymbol{\alpha}^{*}\cdot \boldsymbol{\alpha})
 \right]^{N+n_{0}}\left.\right|_{\gamma_{1}=\gamma_{2}=1}=
 \nonumber \\
 & \sum_{k=n_{0}}^{N+n_{0}}\left(\begin{array}{c}
N+n_{0}\\
k\end{array}\right)
\left[\frac{k!}{(k-n_{0})!}\right]^{2}(\boldsymbol{\alpha}^{*}\cdot
\boldsymbol{\alpha})^{k}
\nonumber \\
 & \frac{{\rm d}^{n_{0}}}{{\rm d}\gamma_{1}^{n_{0}}}\frac{{\rm d}^{n_{0}}}
 {{\rm d}\gamma_{2}^{n_{0}}}\gamma_{1}\gamma_{2}
 \left[1+\gamma_{1}\gamma_{2}(\boldsymbol{\alpha}^{*}\cdot \boldsymbol{\alpha})
 \right]^{N+n_{0}-1}\left.\right|_{\gamma_{1}=\gamma_{2}=1}=
 \nonumber \\
 & \sum_{k=max(n_{0}-1,0)}^{N+n_{0}-1}\left(\begin{array}{c}
N+n_{0}-1\\
k\end{array}
\right)\left[\frac{(k+1)!}{(k+1-n_{0})!}\right]^{2}
\nonumber \\
& \hspace{1cm} \times (\boldsymbol{\alpha}^{*}\cdot \boldsymbol{\alpha})^{k}
\nonumber \\
 & \frac{{\rm d}^{n_{0}}}{{\rm d}\gamma_{1}^{n_{0}}}
 \frac{{\rm d}^{n_{0}}}{{\rm d}\gamma_{2}^{n_{0}}}
 \gamma_{2}\left[1+\gamma_{1}\gamma_{2}(\alpha^{*}\cdot \alpha)\right]^{N+n_{0}-1}
 \left.\right|_{\gamma_{1}=\gamma_{2}=1}=
 \nonumber \\
 & \sum_{k=n_{0}}^{N+n_{0}-1}\left(\begin{array}{c}
N+n_{0}-1\\
k\end{array}\right)
\left[\frac{k!}{(k-n_{0})!}\right]
\frac{(k+1)}{(k+1-n_{0})}
\nonumber \\
& \hspace{1cm} \times (\boldsymbol{\alpha}^{*}\cdot \boldsymbol{\alpha})^{k}
\nonumber 
\end{align}
\begin{align}
 & \frac{{\rm d}^{n_{0}}}{{\rm d}\gamma_{1}^{n_{0}}}
 \frac{{\rm d}^{n_{0}}}{{\rm d}\gamma_{2}^{n_{0}}}(\gamma_{1}\gamma_{2})^{2}
 \left[1+\gamma_{1}\gamma_{2}(\boldsymbol{\alpha}^{*}\cdot \boldsymbol{\alpha})
 \right]^{N+n_{0}-2}\left.\right|_{\gamma_{1}=\gamma_{2}=1}=
 \nonumber \\
 & \sum_{k=max(n_{0}-2,0)}^{N+n_{0}-2}\left(\begin{array}{c}
N+n_{0}-2\\
k\end{array}\right)\left[\frac{(k+2)!}{(k+2-n_{0})!}\right]^{2}
\nonumber \\
& \hspace{1cm} \times (\boldsymbol{\alpha}^{*}\cdot \boldsymbol{\alpha})^{k}
\nonumber \\
 & \frac{{\rm d}^{n_{0}}}{{\rm d}\gamma_{1}^{n_{0}}}
 \frac{{\rm d}^{n_{0}}}{{\rm d}\gamma_{2}^{n_{0}}}(\gamma_{2})^{2}
 \left[1+\gamma_{1}\gamma_{2}(\boldsymbol{\alpha}^{*}\cdot \boldsymbol{\alpha})
 \right]^{N+n_{0}-2}\left.\right|_{\gamma_{1}=\gamma_{2}=1}=
 \nonumber \\
 & \sum_{k=max(n_{0},0)}^{N+n_{0}-2}\left(\begin{array}{c}
N+n_{0}-2\\
k\end{array}\right)
\left[\frac{(k+2)!}{(k+2-n_{0})!}\right]\left[\frac{k!}{(k-n_{0})!}
\right]
\nonumber \\
& \hspace{1cm} \times (\boldsymbol{\alpha}^{*}\cdot \boldsymbol{\alpha})^{k}
~~~.
\label{expand}
\end{align}
For example, the leading term in the first expression is
$\frac{(N+n_{0})!}{N!}n_{0}!$~$(\boldsymbol{\alpha}^{*}\cdot
\boldsymbol{\alpha})^{n_{0}}$ which leads in lowest order in\linebreak
$(\mbox{\boldmath$\alpha$}^{*}\cdot\mbox{\boldmath$\alpha$})$ to the
normalization, as defined in Eq. (8) of Ref. \cite{geom}.  We will,
however assume that $N>>n_{0}>>1$, otherwise the resulting expressions
are too involved.

The equations in (\ref{expand}) can be simplified, using the
abbreviation
\begin{multline}
F_{pq}\left(\alpha^2\right)=\frac{N!^{2}}{(N+n_{0})!}\\
\times \sum_{k={\rm max}(n_{0}-p,n_0-q,0)}^{N+n_{0}-{\rm max}(p,q)}\left(
\begin{array}{c}
N+n_{0}-{\rm max}(p,q) \\
k
\end{array}
\right)
\\
\hspace{1cm} \times \left[\frac{(k+p)!}{(k+p-n_{0})!}\right]
\left[\frac{(k+q)!}{(k+q-n_{0})!}\right]\alpha^{2k}
\end{multline}
which leads then to Eq. (\ref{exp-h}).

\section*{Acknowledgements}

We gratefully acknowledge financial help from DGAPA, from the National
Research Council of Mexico (CONACyT), OTKA (grant No. K72357), and
from the MTA-CONACyT joint project.
Useful discussions with Roelof Bijker (ICN-UNAM)
are acknowledged, related to the definition of the radial distance.
The authors are also thankful to J\'ozsef Cseh for illuminating 
discussions on the subject. 

\end{document}